\documentclass[traditabstract]{aa} 

\usepackage{graphicx}
\usepackage{txfonts}
\usepackage{natbib}
\bibpunct{(}{)}{;}{a}{}{,}

\def\tacc      {$\tau_{acc}$}
\def\tgas      {$\tau_{gas}$}
\def\tdust     {$\tau_{dust}$}
\def\tIRAC     {$\tau_{IRAC}$}
\def\facc      {$f_{acc}$}
\def\fIRAC      {$f_{IRAC}$}

\def\li        {\rm{Li}}
\def\he        {\ion{He}{i}}

\def\li        {\rm{Li}}

\def\kms       {\ts km\ts s$^{-1}$}
\def\msun      {$M_{\odot}$}
\def\myr      {$M_{\odot} yr^{-1}$}

\newcommand{\eg}{{\it e.g.\/}}
\newcommand{\ie}{{\it i.e.\/}}

\begin{document}
\title{Timescale of Mass Accretion in Pre-Main-Sequence Stars}
\titlerunning{Accretion Timescale in PMS stars}
\authorrunning{D. Fedele et al.}
\author{D. Fedele\inst{1},  
M. E. van den Ancker\inst{2},
Th. Henning\inst{1} 
R. Jayawardhana\inst{3}
\and
J. M. Oliveira\inst{4}\fnmsep\thanks{Based on observations collected at
    the European Southern Observatory, Paranal, Chile (Proposal ID:
    078.C-0282; 081.C-0208)}
}

\institute{Max Planck Institut f\"ur Astronomie, K\"onigstuhl 17, D-69117
  Heidelberg, Germany  \\
  \email{dfedele@mpia.de}
\and
  European Southern Observatory, Karl Schwarzschild Strasse 2, D-85748
  Garching bei M\"unchen, Germany 
\and
Department of Astronomy and Astrophysics, University of Toronto, 50 St. George Street, Toronto, ON, M5S 3H4, Canada
\and
Lennard-Jones Laboratories, School of Physical \& Geographical Sciences, Keele
University, Staffordshire ST5 5BG, UK
}
\date{Received ...; accepted ...}

\abstract{We present initial result of a large spectroscopic survey aimed at
  measuring the timescale of mass accretion in young, pre-main-sequence stars
  in the spectral type range K0 -- M5. Using multi-object spectroscopy with
  VIMOS at the VLT we identified the fraction of accreting stars in a number
  of young stellar clusters and associations of ages between 1 -- 50 Myr. The
  fraction of accreting stars decreases from $\sim$ 60~\% at 1.5 -- 2 Myr to
  $\sim$ 2~\% at 10 Myr. No accreting stars are found after 10 Myr at a
  sensitivity limit of $10^{-11}$ \myr. We compared the fraction of stars
  showing ongoing accretion (\facc) to the fraction of stars with near-to-mid
  infrared excess (\fIRAC). In most cases we find \facc ~$<$ \fIRAC ~, \ie,
  mass accretion appears to cease (or drop below detectable level) earlier
  than the dust is dissipated in the inner disk. At 5 Myr, 95~\% of the
  stellar population has stopped accreting material at a rate of $\gtrsim
  10^{-11}$ \myr, while $\sim$ 20~\% of the stars show near-infrared excess
  emission. Assuming an exponential decay, we measure a mass accretion
  timescale (\tacc) of 2.3 Myr, compared to a near-to-mid infrared excess
  timescale (\tIRAC) of 2.9 Myr. Planet formation, and/or migration, in the
  inner disk might be a viable mechanism to halt further accretion onto the
  central star on such a short timescale.}

\keywords{Accretion disks -- Stars : pre-main-sequence -- Planetary systems:
  protoplanetary disks}

\maketitle

  \section{Introduction}
  Circumstellar disks around young, pre-main-sequence stars are the natural loci
  of planet formation \citep[\eg][]{henning08}. Such a {\em protoplanetary}
  disk is formed during the collapse of the molecular cloud to conserve the
  initial angular momentum. It is made of interstellar gas and dust. The
  conventional planet formation model is the so-called {\it core-accretion}
  model \citep[\eg][]{pollack96,mordasini08}. In the simulations of
  \citet{pollack96}, the formation of solar-system like planets takes up to 16
  Myr. Giant planet formation is much faster ($\sim$ 3 Myr) if planet
  migration is included \citep{alibert04,alibert05}. Gravitational instability
  in the disk was also proposed as a viable scenario to form planets,
  especially in the outer disk \citep{boss97, boss98, durisen07}. If
  gravitational instabilities occur, the disk may fragment and form dense
  self-gravitating clumps which are the precursor of gas giant planets. In
  this model planet formation is very fast ($\lesssim$ 1 Myr).
  
  Planet formation models face the fast evolution of protoplanetary
  disks. Star forming regions show a steady decline with time in
  the fraction of stars having infrared excess emission (e.g. \citealp{haisch01,
    haisch05, bouwman06, hillenbrand08}). This is thought to be caused by gradual clearing of
  dust in the inner disk. The growing consensus is that the warm small dust
  grains disappear within the first 5$\div$9 Myr (\tdust). The quantity \tdust ~is
  sometimes adopted as the disk lifetime. Although dust is essential to form
  a planet, in terms of disk mass, and hence dynamics and evolution, it
  accounts only for a small percentage. The bulk of the disk mass is
  thought to be gaseous. At present, it is unclear whether the gas dissipation
  timescale (\tgas) is similar to \tdust.
  
  In this paper we report on a study aimed at determining the timescale for mass
  accretion (\tacc) in protoplanetary disks. The timescale \tacc, i.e. the time at which the
  disk accretion phase ceases, provides a strong constraint on \tgas. Gas may
  still be present after \tacc, however, the amount of remaining gas, and
  hence of disk mass, may be too low to be able to form giant planet(s). Our
  observational strategy, observations and data reduction are presented in
  section 2. Analysis and results are presented in section 3 and 4
  respectively. In section 5 we compare our result with literature
  data. Conclusions are drawn in section 6.

 \begin{table*}
  \caption{Regions observed with VIMOS. The surveyed area for each cluster is
    16\arcmin$\times$18\arcmin  ~centered at the coordinates listed below. The
    limiting magnitude is V = 21 mag. MOS observations were carried out with
    the ``HR-Orange'' grism (maximum spectral coverage: 5200 -- 7600 \AA;
    $\lambda / \Delta\lambda$ = 2150) and a slit width of 1$\arcsec$.}
  \label{tab:obs}
  \centering
  \begin{tabular}{lllllll}
    \hline\hline
    Cluster          & Observing period & RA(J2000) &DEC(J2000)& Distance & A$_V$ & Sources \\
                     & & [hh:mm:ss]&[dd:mm:ss]& [pc]     & [mag] & [N]     \\
    \hline                                                                            
    $\sigma$ Orionis & Oct. 2006 -- Mar. 2007 & 05:38:00 & -02:37:00 & 360     & 0.15  & 216  \\
    NGC 6231         & Apr. 2008 -- Sep. 2008 & 16:54:10 & -41:49:30 & 1240    & 1.36  & 573  \\
    NGC 6531         & Apr. 2008 -- Sep. 2008 & 18:04:13 & -22:29:24 & 1210    & 0.87  & 627  \\
    ASCC 58          & Oct. 2006 -- Mar. 2007 & 10:15:07 & -54:58:12 & 600     & 0.28  & 370  \\
    NGC 2353         & Oct. 2006 -- Mar. 2007 & 07:14:30 & -10:16:00 & 1120    & 0.22  & 316  \\
    Collinder 65     & Oct. 2006 -- Mar. 2007 & 05:26:05 & +15:41:59 & 310     & 0.40  & 198  \\
    NGC 6664         & Apr. 2008 -- Sep. 2008 & 18:36:42 & -08:13:00 & 1440    & 2.20  & 508  \\
    \hline\hline
  \end{tabular}
\end{table*}
 
  \section{Method}\label{sec:method}
  In order to measure \tacc ~in a secure way, we performed an optical
  spectroscopic survey of a large number of stars towards seven young
  stellar clusters (Table \ref{tab:obs}). Multi-object spectroscopy was
  performed with
  VIMOS\footnote{http://www.eso.org/sci/facilities/paranal/instruments/vimos/}
  \citep{lefevre} at the VLT. No a {\em priori} target selection was made, but
  instead we optimized the position of the slits in the multi-object masks in
  order to get the spectrum of as many stars as possible in the proximity of
  the cluster center. We did not aim at taking the spectrum of each (known)
  cluster member, but rather at building an unbiased inventory of the stellar
  population in a sub-region of the cluster down to a limiting magnitude of V
  = 21 mag. For each cluster we estimated the fraction of stars with ongoing
  mass accretion by analyzing the H${\alpha}$ ~line profile. Finally, the
  fraction of accreting stars is plotted against the age of the clusters. In
  this paper we present the analysis of the mass accretion evolution. Results
  on individual clusters will be presented in a forthcoming paper.
  
  The following criteria were adopted for the sample selection: 
  
  \begin{description}
  \item[{\it Age}]: the clusters have ages covering the interval 2 -- 30 Myr; 
  \item[{\it Distance and extinction}]: nearby and optically visible systems
    were selected to ensure detection of low mass stars down to spectral
    type M3 (at least); 
  \item[{\it Stellar density profile}]: to avoid any bias introduced by,
    \eg, mass segregation, we only selected clusters known to have a spherical
    density profile; 
  \item[{\it Angular radius}]: (also linked to previous point) only
    clusters with angular size comparable to the field of view (FOV) of
    VIMOS were considered in order to minimize the number of contaminants.
  \end{description}
    
  For each cluster an area of 16\arcmin$\times$18\arcmin ~was
  investigated acquiring spectra of $\sim$ 200 - 600 representative
  objects in the central region of the cluster. Multiple dithered exposures
  were taken. Details about observations are given in Table \ref{tab:obs}.
  
  Data reduction was performed with IRAF and specific IDL routines. The
  spectra were reduced using a long-slit approach, i.e., extracting each
  spectrum separately. Due to distortion of the instrument night time wavelength
  calibration exposures were taken. No night flat field images were taken. For
  this reason we preferred to not correct the MOS spectra for flat field to
  not introduce further source of noise. The following scheme was applied:
  bias subtraction, wavelength calibration, background removal and spectrum
  extraction using the Horne optimal extraction method \citep{horne}.
  Multiple dithered spectra were combine afterwards using a specific IDL
  routine: before combining the spectra, a scaling factor was applied to take
  into account slit losses in different exposures.
  
\begin{figure*}
    \centering
    \includegraphics[width=17cm]{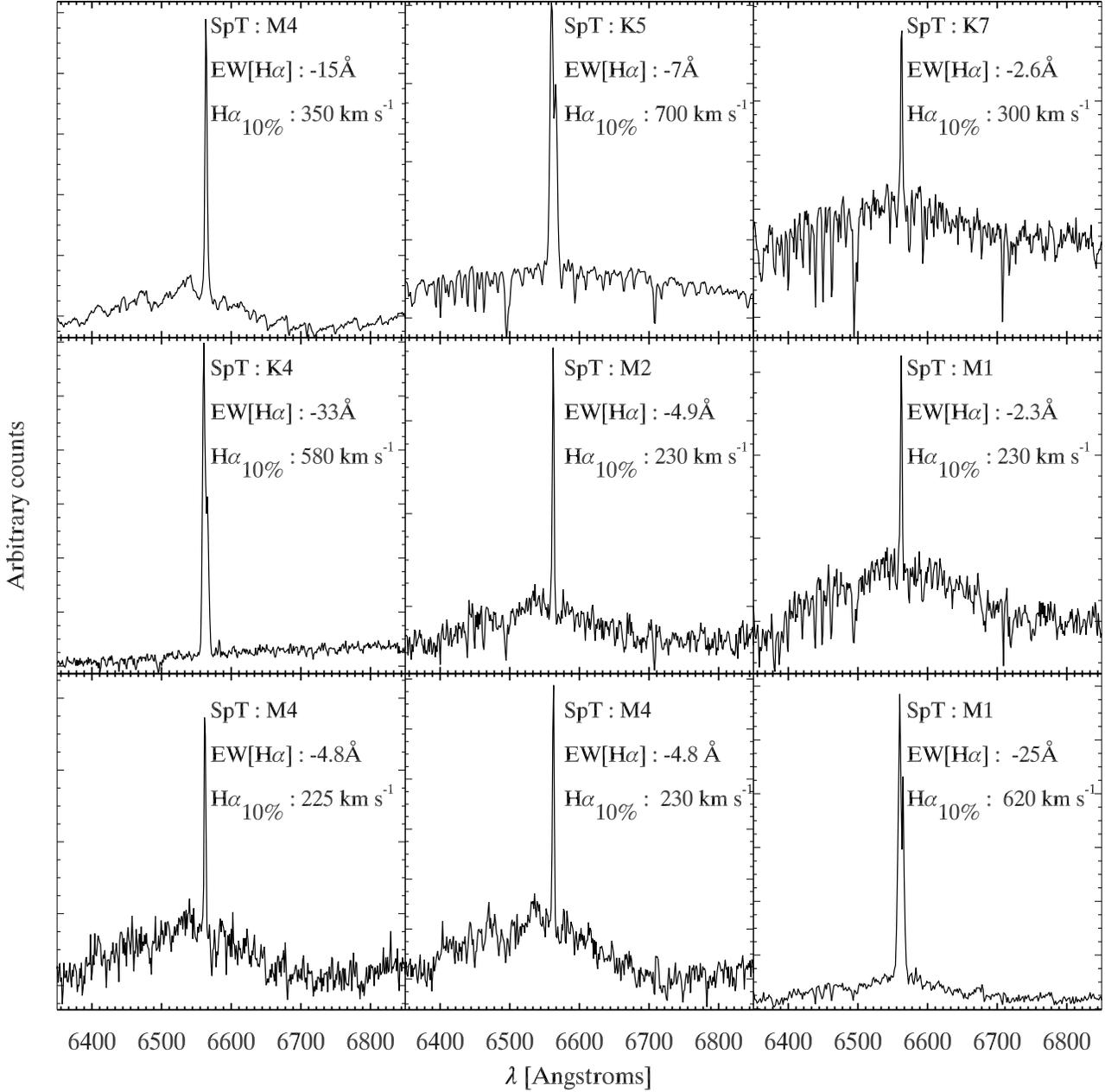}
    \caption{Some examples of VIMOS spectra}
    \label{fig:spec}
  \end{figure*}

  Fig. \ref{fig:spec} shows some examples of VIMOS spectra of our
    program stars.

  \section{Analysis}
  \subsection{Spectral type}
 The spectral type of the objects classified as cluster members was
 derived. For this purpose we used the TiO index defined as:

    \begin{equation}
      TiO (7140\AA) = \frac{C(7020 - 7050 \AA)}{TiO(7125 - 7155 \AA)}
    \end{equation}
    
    where C(7020 -- 7050 \AA) is the spectral continuum computed between 7020
    -- 7050 \AA ~and TiO(7125 -- 7155 \AA) is the intensity of the TiO molecular band
    absorption between 7125 -- 7155 \AA ~\citep{allen95, briceno98,
      oliveira03}. The TiO molecule is very sensitive to the stellar gravity and
    correlates very well with the spectral type in the range K5 -- M6. We
    computed the spectral type of the VIMOS spectra, following the scheme of
    \citet{jeffries07}. For earlier spectral types (K0 -- K5) we directly compared
    the VIMOS spectra with a set of standard stars spectra covering the same
    spectral range and having similar spectral resolution. The uncertainty on
    spectral type derived with these techniques is one sub-class.
    
    In the following analysis we consider only stars having spectral type between
    K0 and M5. 
  
  \subsection{Membership}
  The cluster membership was established on the basis of the presence of two
  spectral features: emission of the H${\alpha}$ line at 6563 \AA ~and
  absorption of the \li ~6708 \AA ~line. H${\alpha}$ emission in young stars
  is either produced by mass accretion or by chromospheric activity. The
  presence of \li ~is also widely used as youth indicator
  \citep[\eg][]{basri91}; \li ~is depleted very fast once the thermonuclear
  reactions begin in the core of the star. The abundance of \li ~in the
  stellar atmosphere is inversely proportional to the age of the star
  \citep{jeffries07,mentuch08}. This youth indicator is best suited for K- and
  M-type stars younger than $\sim$ 30 Myr \citep{jeffries07}. Some stars have
  strong \li ~absorption in their spectrum but no H${\alpha}$ emission. These
  might be cluster members with no, or reduced, chromospheric activity. We
  measured the EW of \li ~6708 \AA ~of these sources and compared this with the
  typical \li ~6708 \AA ~EW measured for the certain members of the cluster
  (0.2 \AA ~$<$ EW [\li] ~$<$ 0.6 \AA). We note that the lower limit of EW
  Li~measured here (0.2 \AA) is an intermediate value assumed as Li~indicator
  applied by other authors (\eg ~\citealt{jeffries07}, EW $>$ 0.3 \AA;
  \citealt{winston09}, EW $>$ 0.1 \AA). In most of the cases the spectra show
  either a tiny (EW $\lesssim$ 0.1 \AA) H${\alpha}$ emission or small
  absorption. In the latter case, the absorption is much smaller than the
  typical photospheric absorption for the same spectral type. The ``reduced''
  H${\alpha}$ absorption might result from {\it line veiling} caused by the
  chromospheric activity. These stars are classified as cluster members.

  \subsection{Signature of mass accretion}\label{subsec:macc}
  To distinguish between chromospheric and accretion origin of the H${\alpha}$
  emission, \citet{white03} suggest a different threshold in EW for different
  spectral types. \citet{barrado03} have defined an empirical criterion based
  on the saturation limit of the chromospheric activity. Another tool of
  investigation is the width of H${\alpha}$ at 10\% (H$\alpha_{10\%}$) of the
  line peak \citep{white03, natta04, jay06, flaherty08}. As shown in Fig. 7 of
  \citet{white03}, accreting objects have systematically larger value of
  H$\alpha_{10\%}$ compared to non-accreting objects. Stars with
  H$\alpha_{10\%}$ $>$ 270 \kms ~are accreting material from a circumstellar
  disk. This criterion is valid down to a mass accretion rate of $10^{-10}
  \div 10^{-12}$ \myr~ \citep{natta04}. The actual limit on detectable
  accretion rate depends on the stellar mass and age. As a conservative value
  we adopt 10$^{-11}$ \myr. In this work we used both EW [H${\alpha}$] and
  H$\alpha_{10\%}$ criterion to distinguish between accretion and
  chromospheric activity. This method is solid to find signature of accretion
  whenever EW [H${\alpha}$] and H$\alpha_{10\%}$ are above, or close to the
  thresholds \citep{jay06}. However, stars with values lower than the cutoff
  values may still be accreting at low rate ($< 10^{-11}$ \myr) and cannot be
  distinguished from pure chromospheric activity. 
  
  The spectral resolution of VIMOS is enough to measure the widening of
  the H$\alpha$ line due to accetion. At 6563 \AA ~the minimum velocity
  resolvable with VIMOS is 140 \kms.
  We measured the EW and the 10\% width of the H${\alpha}$ emission. The
  spectral continuum was computed using two spectral windows at both sides of
  the H${\alpha}$ line. The continuum level in correspondence of the emission
  line was interpolated using a first order polynomial. The spectra were then
  continuum-normalized and subtracted. H$\alpha_{10\%}$ was measured as the
  width of the H${\alpha}$ line at the height corresponding to 10\% of the
  line peak. In the case of narrow emission line and low signal-to-noise
  spectrum, the 10\% level might be confused with the neighbor continuum
  fluctuations. To avoid this, the result was visually inspected. The major
  source of uncertainty for the measurements of EW [H${\alpha}$] and
  H$\alpha_{10\%}$ is the determination of the underlying spectral
  continuum. The errors on H${\alpha}$ EW and H$\alpha_{10\%}$ were computed
  by taking the continuum error into account. For most of the spectra the
  signal-to-noise ratio close the H$\alpha$ line is $>$ 10 - 20. This
  translate in a precision of $\sim$ 0.1 \AA ~and 5 \kms ~for equivalent width
  and H$\alpha_{10\%}$, respectively. A further contribution of uncertainty is
  given by the instrument's spectral precision. We determined this by
  computing the full widht half maximum and standard deviation for a number of
  arc lines in proximity of the H${\alpha}$ line. The standard deviation is of
  the order of 5 -- 7\kms. We therefore assume 0.2\AA ~and 10\kms ~as a
  (conservative) lower limit to the uncertainty of EW [H${\alpha}$] and
  H$\alpha_{10\%}$, respectively.
  
  \begin{figure*}
    \centering
    \includegraphics[width=9cm, angle=90]{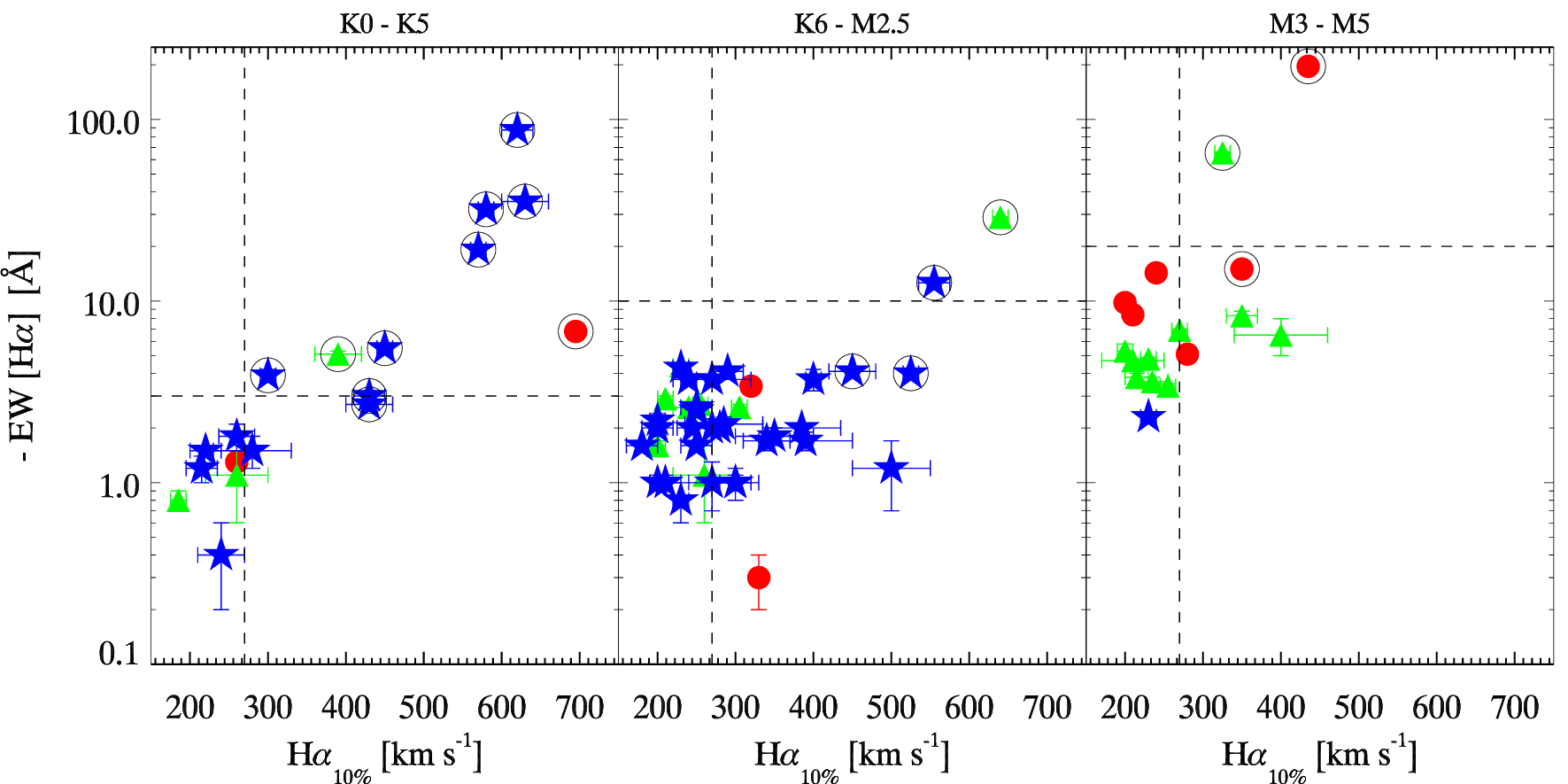}
    \caption{EW [H$\alpha$] vs H$\alpha_{10\%}$ for three different spectral
      type ranges. Vertical and horizontal dashed lines represent the
      thresholds for accreting stars. The different symbols refer to different
      clusters : dots -- Sigma Ori, triangles -- NGC 6531, stars -- NGC
      6231. Objects classified as accreting stars are over plotted with
        open circles.}
    \label{fig:ewha10}
  \end{figure*}

  \section{Result}
  None of the stars in regions older than 10 Myr in our survey (ASCC 58,
  Collinder 65, NGC 2353 and NGC 6664) shows evidence of ongoing accretion
  (Table \ref{tab:result}). Results for the youngest ($<$ 10 Myr) regions are
  described below. In  Fig. \ref{fig:ewha10} we plot EW [H${\alpha}$] versus
  H$\alpha_{10\%}$ for three different spectral ranges. Vertical and
  horizontal dashed lines represent the thresholds for accreting stars. The
  different symbols refer to different clusters : dots -- $\sigma$ Ori, triangles
  -- NGC 6531, stars -- NGC 6231.

  As shown in Fig. \ref{fig:ewha10} there are a number of sources with EW
  [H${\alpha}$] below the accretion threshold but with H$\alpha_{10\%}$
  $>$ 270 \kms. If the line broadening is caused by mass accretion at a rate
  $> 10^{-11}$ \myr, we would expect to measure a large EW as well. Beside
  accretion, stellar binarity and fast rotation might also be responsible
  for the line broadening. In order to decipher the nature of these sources we
  also investigated the presence of other emission lines which are
  associated with mass accretion \citep[e.g. \he~5876\AA, 6678\AA;
    e.g.][]{beristain01, herczeg08}. In the presence of these accretion
  diagnostics, we can safely state that the star is accreting.
  
  \subsubsection*{$\sigma$ Ori}
  10 out of the 216 sources with spectra taken with VIMOS are classified as
  members of the $\sigma Ori$ cluster. All of them show emission in the
  H${\alpha}$ line as well as presence of Li. 2 out of 10 objects are clearly
  accreting stars according to the method mentioned above. A further object
  ($\sigma$ Ori \# 1: RA = 84.584154, DEC = -2.633792, Sp. Type M5) has a
  broad H$\alpha_{10\%}$ (350 \kms) but an H${\alpha}$ EW value (15 \AA) below
  the accreting threshold. The presence of \he emission lines (~5876 \AA
  ~EW = -0.75, ~6678 \AA ~EW = -0.2) suggests the presence of hot gas in the
  vicinity of the star and hence signature of ongoing accretion 
    \citep{beristain01}. The fraction of accreting objects in $\sigma$ Ori
  derived here is 3/10 or 30 \% $\pm$ 17 \%\footnote{computed as
      $\sqrt{N_{*, ~acc}}/N_{*,total}$}. This is in agreement with previous
  result of \citet[][30--40 \%]{zapatero02}, \citet[][27 $\pm$ 7
    \%]{barrado03}, \citet[][30 \%]{oliveira06}. The good agreement with
  previous results validates our observational strategy despite the low
    number statistics.

  \subsubsection*{NGC6231}
  Out of 573 sources in the region of NGC 6231 we identified 78
  objects as cluster members. The majority of these sources have spectral
  type in the range K0 -- M5 (there are 3 sources of earlier spectral type)
  and show both H${\alpha}$ emission and strong Li ~absorption. 11 out
  of 78 objects (all late K type) were identified based only on the
  presence of Li. Each of these spectra was compared with a standard star of
  equal spectral type. All these spectra show either a tiny (EW $\sim$ 0.1
  \AA) H${\alpha}$ emission or small absoption. In the latter case, the
  absoption is much smaller than the typical photospheric absorption for the
  same spectral type. Further 3 objects have no H${\alpha}$ and the presence
  of Li 6707.8 absorption is not clear. 7 out of the 75 K0 -- M5 sources
  in NGC 6231 are consistent with ongoing mass accretion with both
  diagnostics. Further 2 sources have EW [H$\alpha$] close to the
  accretion threshold but large ($>$ 270 \kms) H$\alpha_{10\%}$ (panel K0 --
  K5 in Fig. \ref{fig:ewha10}). Many stars with spectral type in the range
  K6 -- M2.5 have large H$\alpha_{10\%}$ but small EW [H$\alpha$]. We
  inspected all of them and found 2 objects with spectral type K7 showing
  \he~5876\AA ~in emission (EW = -0.5\AA, -0.6\AA ~respectively). The evidence
  of large H$\alpha_{10\%}$ together with the \he ~emission is most
  likely due to ongoing mass accretion and these two stars are classified as
  accreting stars. We estimate a fraction of accreting stars in NGC 6231 of
  11/75 or 15 ($\pm$ 5 \%). We warn the reader that this might be a lower
  limit to the actual fraction of accreting stars; further investigation is
  needed to disentagle the nature (accretion vs binarity/rapid rotation) of
  the systems with large H$\alpha_{10\%}$ ($>$ 300 \kms) but low EW
  [H$\alpha$].
  
  \subsubsection*{NGC6531}
  We identified 26 cluster members in NGC 6531 based on the presence of
  H${\alpha}$ emission and presence of Li. Further 13 sources show presence
  of Li ~6708 \AA, but have H${\alpha}$ in absorption. As in the case of NGC
  6231, these might be cluster members with no, or reduced, chromospheric
  activity. We measured the EW of Li ~6708 \AA~ of these 13 sources and
  compared with the typical EW of the 26 stars in NGC 6531 showing also
  H${\alpha}$ emission (0.2 \AA~ $<$ EW Li ~$<$ 0.6 \AA). 10 of the 13 sources
  have similar EW and are likely cluster members. The remaining 3 sources have
  EW between 0.1 -- 0.2 \AA. Moreover, these are characterized by strong
  absorption lines at 5778, 5796, 6284, 6614 \AA~ which are produced by
  diffuse interstellar absorption bands (DIBs). These are likely located in
  the background of the cluster and are classified as non-members. Out of the
  36 identified members, 3 objects show broad and strong H${\alpha}$ emission
  consistent with mass accretion. One further object is classified as
  uncertain accreting star. The fraction of accreting stars is then 8 $\pm$
  5\%. Given the large H$\alpha_{10\%}$ ($>$ 300 \kms) of 2 objects in the
  spectral range M3 -- M5, the fraction of accreting stars computed here might
  be a lower limit. Follow up observations are needed for these two sources.

  \section{Discussion}
\begin{figure}
    \centering
    \includegraphics[width=9cm]{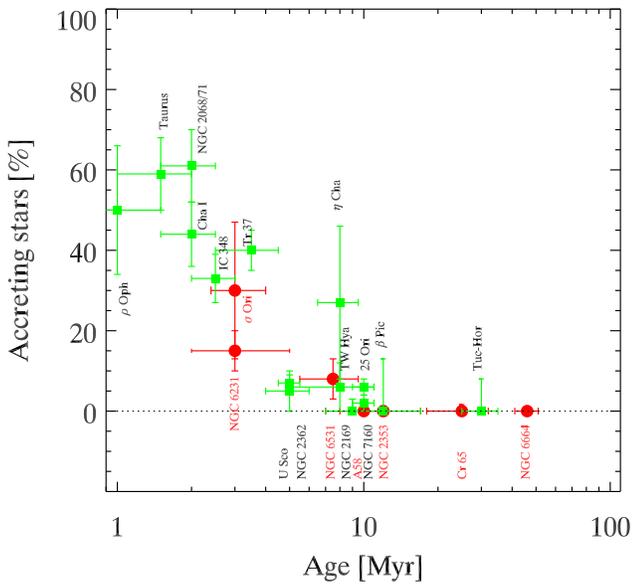}
    \caption{Accreting stars frequency as a function of age. New data (based on
      the VIMOS survey) are shown as (red) dots, literature data as (green)
      squares. Colored version is available in the electronic form.}
    \label{fig:frac}
  \end{figure}

  The fraction of accreting stars (\facc) for each cluster is listed in Table
  \ref{tab:result} and plotted in Fig. \ref{fig:frac} together with literature
  data. 

  The aim of this paper is to trace the evolution of mass accretion with
  time. A critical point is the age of the clusters. We adopted the most
  recent age measurements (Table \ref{tab:result}). For NGC 6231
  \citet{sana07} find that the bulk of the CTTS has an age between 2 -- 4 Myr,
  although with a large age spread. The most recent study of $\sigma$ Ori
  suggests an age of 3 Myr \citep[\eg][]{caballero08}. For NGC 6531
  \citet{park01} measure an age of 7.5 $\pm$ 2 Myr in agreement with
  \citep{forbes} who measure 8 $\pm$ 3 Myr. For the remaining clusters we
  adopt the age provided by WEBDA\footnote{http://www.univie.ac.at/webda/,
    operated at the Institute for Astronomy of the University of Vienna}. For
  these clusters the age is estimated by fitting the zero-age-main-sequence
  (ZAMS) to the brightest stars of the cluster. We take an accuracy of 30\% as
  a conservative assumption.

  In Fig. \ref{fig:frac} the fraction of accreting stars as function
  of cluster age is shown. The results obtained with VIMOS are shown as filled
  circles. Measurements of \facc ~exist in literature for some clusters. These
  are are shown as filled squares in Fig. \ref{fig:frac} and are listed in
  Table \ref{tab:result}. We considered only determinations of \facc, using
  the method of \citet{white03} and/or \citet{barrado03}. The fraction of
  accreting stars may vary with spectral type. In the case of Upper Sco
  \citet{mohanty05} \eg, found an increase from 7\% ($\pm$ 2\%) for spectral
  type earlier than M4 to 20\% ($\pm$ 10\%) for later spectral types. For the
  youngest regions, $\rho$ Oph, Taurus, Cha I and IC 348, \citet{mohanty05}
  find no appreciable variation, within errors, in the fraction of accreting
  sources between the two mass ranges. For consistency with our survey we
  include here only the results for spectral types between K0 -- M5.

  \begin{table*}
    \caption{Adopted age, spectral type range, \facc ~and \fIRAC ~(when
      available) in Fig. \ref{fig:frac} and \ref{fig:comparison}. References:
      \citet[S82]{schimdt82}, \citet[P01]{park01}, \citet[Ha05]{hartmann05},
      \citet[K05]{kharchenko05}, \citet[M05]{mohanty05},
      \citet[SA05]{sicilia05}, \citet[C06]{carpenter06}, \citet[L06]{lada06},
      \citet[JA06]{jay06}, \citet[SA06]{sicilia06}, \citet[D07]{dahm07},
      \citet[B07]{briceno07}, \citet[JE07]{jeffries07},
      \citet[He07]{hernandez07}, \citet[S07]{sana07},
      \citet[C08]{caballero08}, \citet[FM08]{flaherty08},
      \citet[L08]{luhman08}, \citet[S09]{sicilia09},
      \citet[ZS04]{zs04}} \label{tab:result}
    \centering
    \begin{tabular}{llllllll}
      \hline\hline
      Cluster      & Age    & Sp.T range & \facc       & \fIRAC    & Age ref. & \facc ~ref. &  \fIRAC ~ref.\\
                   & [Myr]  &            &[\%]         & [\%]     &          &             &\\
      \hline                                                                
      $rho$ Oph    &   1    & K0 -- M4   & 50 $\pm$ 16 &          & M05  & M05      &      \\
      Taurus       &   1.5  & K0 -- M4   & 59 $\pm$ 9  & 62       & M05  & M05      & Ha05 \\
      NGC 2068/71  &   2    & K1 -- M5   & 61 $\pm$ 9  & 70       & FM08 & FM08     & FM08 \\
      Cha I        &   2    & K0 -- M4   & 44 $\pm$ 8  & 52 -- 64 & Lu08 & M05      & Lu08 \\
      IC348        &   2.5  & K0 -- M4   & 33 $\pm$ 6  & 47       & L06  & M05      & L06  \\
      NGC 6231     &   3    & K0 -- M3   & 15 $\pm$ 5  &          & S07  & this work&      \\
      $\sigma$ Ori &   3    & K4 -- M5   & 30 $\pm$ 17 & 35       & C08  & this work& He07 \\
      Trumpler 37  &   3.5  & K0 -- M3   & 40 $\pm$ 5  & 47       & SA06 & SA06     & SA06 \\
      Upper Sco    &   5    & K0 -- M4   & 7  $\pm$ 2  & 19       & C06  & M05      & C06  \\
      NGC 2362     &   5    & K1 -- M4   & 5  $\pm$ 5  & 19       & D07  & D07      & D07  \\
      NGC 6531     &   7.5  & K4 -- M4   & 8  $\pm$ 5  &          & P01  & this work&      \\
      $\eta$ Cha   &   8    & K4 -- M4   & 27 $\pm$ 19 & 50       & S09  & JA06     & S09  \\
      TWA          &   8    & K3 -- M5   & 6  $\pm$ 6  &          & D06  & JA06     &      \\
      NGC 2169     &   9    & K5 -- M6   & 0$^{+3}$    &          & JE07 & JE07     &      \\
      25 Ori       &   10   & K2 -- M5   & 6  $\pm$ 2  &          & B07  & B07      &      \\   
      NGC 7160     &   10   & K0 -- M1   & 2  $\pm$ 2  & 4        & SA06 & SA05     & SA06 \\      
      ASCC 58      &   10   & K0 -- M5   & 0$^{+5}$    &          & K05  & this work&      \\
      $\beta$ Pic  &   12   & K6 -- M4   & 0$^{+13}$   &          & ZS04 & JA06     &      \\
      NGC 2353     &   12   & K0 -- M4   & 0$^{+6}$    &          & K05  & this work &      \\
      Collinder 65 &   25   & K0 -- M5   & 0$^{+7}$    &          & K05  & this work &      \\
      Tuc-Hor      &   27   & K1 -- M3   & 0$^{+8}$    &          & ZS04 & JA06      &      \\
      NGC 6664     &   46   & K0 -- M1   & 0$^{+4}$    &          & S82  & this work &      \\
      \hline\hline
    \end{tabular}
  \end{table*}

  \subsection{Evolution of \facc}
  The fraction of accreting stars decreases quickly with time within the first
  10 Myr. There is a clear trend from the 1.5 -- 2 Myr old regions (\facc ~=
  60\%) down to the 10 Myr old clusters (mean \facc~= 2\%). At the age of 5 Myr
  the mean \facc~is $\sim$ 5\% drastically lower than the 3--4 Myr old regions
  of $\sigma$ Ori, NGC 6231 and Trumpler 37 (average 28\%). No
  accreting stars are found beyond 10 Myr. This is in good agreement with
  previous measurements of mass accretion timescale by
  \citet[\eg][]{mohanty05, jay06}. There are some outliers. The $\eta$ Cha
  cluster shows an higher frequency of accretors compared to clusters of equal
  age \citep[\eg][]{sicilia09}. This particular association shows a similar
  behavior in the disk frequency (from near-infrared excess). Age and distance
  are unlikely to be wrong for this well know system. \citet{moraux07} suggest
  that most of the low-mass members of the association have been dispersed by
  dynamical evolution. The current list of members of $\eta$ Cha might be
  biased towards infrared-excess/H${\alpha}$ emitting sources. Inversely the
  very young region of $\rho$ Oph lies below the trend. The statistics in this
  region are limited by extinction which may obscure part of the stellar
  population \citep[\eg][]{mohanty05}. Moreover, the result of
  \citet{mohanty05} might be contaminated by an older population of stars;
  objects in the periphery of $\rho$ Oph core appear older than stars in the
  core itself \citep{wilking05}. In the following analysis we will not include
  $\rho$ Oph and $\eta$ Cha.
  
  NGC 6231 (2 -- 4 Myr) also appears to have a low \facc ~for its
  age. Interestingly, this region hosts many O and B stars. External
  photoevaporation (by means of the O, B stars) may accelerate disk
  dissipation. 
  
  \subsection{\facc ~vs \fIRAC}
  In this section we compare the fraction of accreting stars to the
  fraction of stars with near-to-mid infrared excess (\fIRAC). The latter is
  measured as the fraction of stars having infrared excess in the Spitzer/IRAC
  $[3.6]$ -- $[8.0]$ bands over the total fraction of stars in the cluster. 
  The definition of an excess source is based on the slope ($\alpha$) of the
  infrared spectral energy distribution determined from the IRAC $[3.6]$ --
  $[8.0]$ bands \citep{lada06}:

    \begin{equation}
      \alpha_{IRAC} = \frac{d (log(\lambda F_{\lambda}))}{d (log(\lambda))}; \qquad 3.6
      \mu m < \lambda < 8.0 \mu m
    \end{equation}

    Stars with $-1.8 < \alpha < 0$ are classified as class II
    \citep[\eg][]{hernandez07}. In order to coherently compare \fIRAC ~with
    \facc~ we refers only to stars of spectral type between K0 -- M5. Whenever
    \fIRAC ~was measured as function of stellar mass rather than spectral type
    we only consider stars of mass between 0.1 -- 1.2 \msun. By applying these
    selection criteria we find the following results for \fIRAC ~(Table
    \ref{tab:result}): Taurus -- 62\% \citep{hartmann05}, NGC 2068/71 -- 70\%
    \citep{flaherty08}, Cha I -- 52\% $\div$ 64\% \citep{luhman08}, IC 348 --
    47\% \citep{lada06}, $\sigma$ Ori -- 35\% \citep{hernandez07}, Trumpler 37
    -- 47\% \citep{sicilia06}, Upper Sco -- 19\% \citep{carpenter06}, NGC 2362
    -- 19\% \citep{dahm07}, $\eta$ Cha -- 50\% \citep{sicilia09} and NGC 7160
    -- 4\% \citep{sicilia06}.
    
    \smallskip
    \noindent

    Interestingly for some clusters we find a lower fractional value of stars
    with evidence of ongoing accretion compared to stars with near-to-mid
    infrared excess. For example among the 64 K0 -- M5 stars identified by
    \citet{flaherty08} in NGC 2068/71, 39 (61\%) of these show clear signature
    of mass accretion while 45 (70\%) stars have IRAC excess. We note that 2
    of the non accreting stars in their sample are identified by strong IRAC
    excess \citep[ID 416, 843 in][]{flaherty08}, while one transitional object
    (ID 281) is a strong accretor. Similarly, for $\sigma$ Ori we measure an
    \facc ~= 30\% while \fIRAC ~= 35\%. This is consistent with the result of
    \citet{damjanov07} in Chameleon I. They find a small population of
    non-accreting objects bearing a dusty inner disk.
   
    \smallskip
    \noindent

\begin{figure}
  \centering
  \includegraphics[width=9cm]{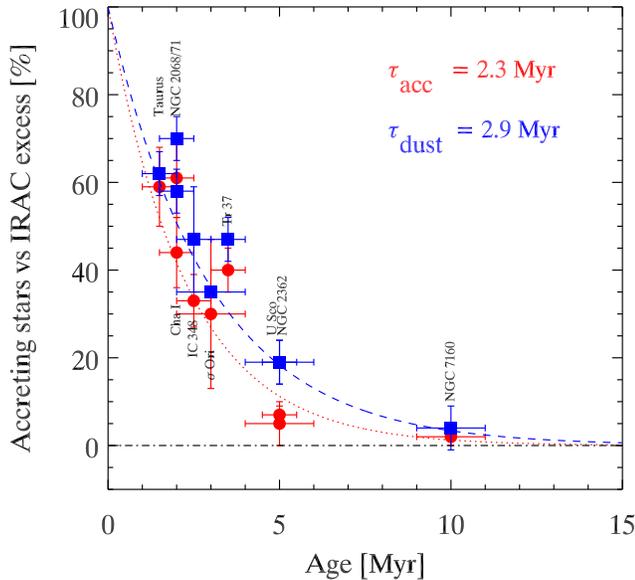}
  \caption{\facc ~(dots) versus \fIRAC ~(squares) and exponential fit (dotted
    line for \facc, dashed line for \fIRAC.}\label{fig:comparison}
\end{figure}

    In Fig. \ref{fig:comparison} we plot \facc ~vs \fIRAC ~for the regions
    where both quantities are known. Assuming an exponential decay we fit
    \facc ~and \fIRAC~ with the following function: 
  
    \begin{equation}
      f_i = C \cdot exp(-t~/~\tau_i) 
    \end{equation}
    
    where C is a constant and it is normalized assuming \facc ~= \fIRAC ~=
    100\% for $t=0$. The fit gives an accretion timescale of \tacc ~= 2.3 Myr
    and a near-to-mid infrared excess timescale of \tIRAC ~= 2.9 Myr.

  \subsection{Inner disk dissipation and planet formation}
  Mass accretion and dust dispersion in the inner disk appear to have different
  timescales. At an age of 5 Myr, 95\% of the total stellar population
  has stopped accreting material at a rate $\gtrsim 10^{-11}$ \myr ~while $\sim$
  20\% of the disks still retain enough dust to produce detectable infrared
  excess. In order to sustain mass accretion at a rate larger than $10^{-11}$
  \myr~ a gas disk reservoir is required. This implies that at an age of \tacc
  ~= 2.3 Myr most of the disk mass is drastically reduced while at an age of 5
  Myr only 5\% of the disks have enough gas mass to keep on accreting material
  onto the central star. Models of viscous evolution
  \citep[e.g.][]{hartmann98} predict a steady decrease of the mass accretion
  rate with time and this might be an explanation to the bimodal evolution of
  \facc ~and \fIRAC. However, the timescale for viscous evolution to stop
  accretion ($>$ 10 Myr) is much longer than our estimate of \tacc ~(2.3
  Myr). The efficiency of viscous evolution might be enhanced if
  photoevaporation is also taken into account as first argued by
  \citet{clarke01}. Their so called ``UV-switch'' model is very efficient in
  stopping further accretion as the accretion rate drops below the
  photoevaporation rate. The advantage of this is the shorter accretion
  timescale. But this model also has a drawback, namely that disk dissipation
  following the end of the accretion phase should occur on a viscous timescale
  ($10^5$ yr). This is in contrast with the relatively high number of stars
  still retaining their dusty disk (e.g. 20\% at 5 Myr) after accretion
  ceases.

  \smallskip
  \noindent

  Another possible explanation is that accretion onto the central
  star is stopped by planet formation, and migration, in the inner disk. If a
  giant planet forms/migrates to the inner region of the disk, it may stop
  further accretion onto the star. 
  In this regard, recent self-consistent numerical simulations of planet
  formation by, e.g., \citet{thommes} are consistent with this scenario.
  In these simulations planetesimal formation occurs over a large
  range of disk radii. As soon as the cores of giant planets form they migrate
  to small radii ($<$ 1 AU). This happens very fast, within 1 -- 2 Myr. The
  disk surface density drops to zero, and so does accretion onto the star,
  within the planet's orbit. Dust grains beyond the orbit of the planet
  instead, might still absorb and re-emit the stellar light and produce the
  measured IRAC excess.

\section{Conclusion}
In this paper we presented an analysis of the evolution of mass accretion in
PMSs in the spectral type range K0 -- M5. These are the main results: (1)
the fraction of stars with ongoing mass accretion decreases fast with time,
going from $\sim$ 60\% at 1.5 -- 2 Myr down to $\sim$ 2\% at 10 Myr; (2) this
fraction is systematically lower than the fraction of stars showing
near-to-mid infrared excess; (3) mass accretion and dust dissipation in the
inner disk appear to have different characteristic timescales, 2.3 and 2.9 Myr
respectively; (4) within 5 Myr the mass accretion rate of 95\% of the stellar
population drops below our detection limit of $10^{-11}$ \myr. While viscous
evolution and photoevaporation might be unable to slow down accretion (and
leave a substantial dust mass) on such a short timescale, planet formation,
and/or migration, in the inner disk (few AU) might be a viable mechanism to
halt further accretion onto the central star.

\begin{acknowledgements}
We thanks the ESO staff for performing the VIMOS observations in service
mode. This research has made use of the SIMBAD database, operated at CDS,
Strasbourg, France. We are grateful to the anonymous referee for his
comments and suggestions.
\end{acknowledgements}


\bibliographystyle{aa} 
\bibliography{fedele09}

\end{document}